# Layer-Wise Security Framework and Analysis for the Quantum Internet

Zebo Yang, *Student Member, IEEE*, Ali Ghubaish, *Student Member, IEEE*, Raj Jain, *Life Fellow, IEEE,* Ala Al-Fuqaha, *Senior Member, IEEE,* Aiman Erbad, *Senior Member, IEEE,* Ramana Kompella, *Member, IEEE,* Hassan Shapourian, *Member, IEEE,* and Reza Nejabati, *Member, IEEE*

*Abstract*—With its significant security potential, the quantum internet is poised to revolutionize technologies like cryptography and communications. Although it boasts enhanced security over traditional networks, the quantum internet still encounters unique security challenges essential for safeguarding its Confidentiality, Integrity, and Availability (CIA). This study explores these challenges by analyzing the vulnerabilities and the corresponding mitigation strategies across different layers of the quantum internet, including physical, link, network, and application layers. We assess the severity of potential attacks, evaluate the expected effectiveness of mitigation strategies, and identify vulnerabilities within diverse network configurations, integrating both classical and quantum approaches. Our research highlights the dynamic nature of these security issues and emphasizes the necessity for adaptive security measures. The findings underline the need for ongoing research into the security dimension of the quantum internet to ensure its robustness, encourage its adoption, and maximize its impact on society.

## I. INTRODUCTION

PROGRESS in quantum technologies is reshaping the methods by which computers communicate and process data, laying the foundation for the emergence of the quantum internet. For instance, recent experiments have demonstrated deterministic delivery of correlated photons between quantum memories at separate locations, utilizing telecommunication wavelengths to extend the transmission range [1], [2]. Satellite-based quantum communication has also emerged as a promising solution for long-distance quantum information transfer, extending the reach of experiments to thousands of kilometers or more [3]. These developments set the stage for a global quantum communication network that could revolutionize secure data transmission.

Unlike classical networks, which utilize bits represented by electrical signals or electromagnetic waves, the quantum internet employs quantum bits (qubits) that can simultaneously exist in a superposition of 0 and 1 states. This feature, along with entanglement, enables certain types of computation that surpass the capabilities of classical computers, such as factorization and unstructured search [4]. Entanglement refers to qubits that are correlated in such a way that the state of one can instantly impact the state of another, regardless of distance. Furthermore, the fact that any measurement on qubits disrupts their current states provides what is known as "unconditional security." These properties will allow the quantum internet to enhance security and computational capabilities. One of its most notable applications is Quantum Key Distribution (QKD) [5], a secure protocol that generates and distributes secret keys such that any eavesdropping attempt is detectable. This is particularly valuable for sensitive communications in sectors such as finance, healthcare, and national security [6].

Beyond security, the quantum internet opens up new fields of application, including distributed quantum computing [7], quantum sensing [8], and private quantum clouds. Distributed quantum computing connects multiple quantum computers to share resources, significantly expanding computational power and supporting solutions for quantum data centers. Quantum sensing provides highly sensitive measurements that advance fields such as medical imaging, navigation, and environmental monitoring. Connecting quantum sensors in a network allows for the development of even more robust and sensitive systems. Quantum private clouds allow cloud computations to remain confidential, even from the cloud provider.

However, the framework for quantum internet architecture remains unsettled, outlining two dominant models [9], [10], [11]. The first model aligns with traditional networking practices, emphasizing point-to-point data transfer through direct qubit transmission over connected channels [12], [13]; we refer to this as a qubit-forwarding network. The second envisions an overlay network that distributes entanglement via intermediary nodes, allowing endpoints to use the final end-to-end entanglement for communication [14], [15]; this is referred to as an entanglement distribution network.

It is worth noting that, as QKD is a prominent application of the quantum internet, some literature uses the term "quantum network" to refer specifically to networks consisting of QKD channels, which are dedicated networks focused on distributing secret keys between endpoints. This paper also considers these

---

˗This work was supported in part by Cisco University Research Grant #92627757 and the Qatar Research, Development, and Innovation (QRDI) Academic Research Grant #ARG01-0501-230053. The statements made here are solely the responsibility of the authors. *(Corresponding author: Zebo Yang)*.

Zebo Yang, Ali Ghubaish, and Raj Jain are with the Department of Computer Science and Engineering, Washington University in St. Louis, St. Louis, MO USA (zebo@wustl.edu; aghubaish@wustl.edu; jain@wustl.edu).

Ala Al-Fuqaha is with the College of Science and Engineering, Hamad Bin Khalifa University in Education City, Doha, Qatar (aalfuqaha@hbku.edu.qa).

Aiman Erbad is with the College of Engineering, Qatar University, Doha, Qatar (aerbad@qu.edu.qa).

Ramana Kompella, Hassan Shapourian, and Reza Nejabati are with Cisco Research, San Jose, CA, USA (rkompell@cisco.com; hassan.shapp@gmail.com; rnejabat@cisco.com).

\#

networks, referred to as QKD networks. In such networks, specialized relays are designed to forward secret keys generated through QKD [16], which enable classical encryption between remote endpoints.

Regardless of network type, there is a consensus that the quantum internet will operate as a hybrid quantum-classical network, integrating components of the existing classical internet infrastructure. It is intended to optimize performance and security, particularly during the Noisy Intermediate-Scale Quantum (NISQ) era, where quantum systems remain sensitive to environmental factors and are limited by scale [17]. Nonetheless, the quantum internet remains in its developmental stages and faces numerous unique security challenges [18]. Key properties of quantum networks—such as superposition, entanglement, and the no-cloning theorem [19]—introduce novel vulnerabilities that traditional security measures cannot effectively address. For example, the no-cloning theorem, which prevents the exact copying of arbitrary unknown quantum states, introduces a potential vulnerability in data recovery and redundancy mechanisms, compared to classical systems that depend on data replication. Mitigating these vulnerabilities will require innovative security paradigms tailored specifically for quantum networks, capable of managing both the inherent uncertainties of quantum systems and their integration with classical infrastructure.

Additionally, while QKD effectively mitigates conventional Man-in-the-Middle (MiTM) attacks, it remains vulnerable to attack strategies that exploit quantum properties, such as Photon Number Splitting (PNS) [20], which compromises the security of weak coherent pulses in QKD systems, as well as time-shift attacks [21], which manipulates timing information to gain an advantage in key extraction. Denial of Service (DoS) attacks further threaten it by disrupting quantum channels or overwhelming system resources [22]. Beyond QKD, the broader network infrastructure faces new security challenges. For example, current quantum network routing relies heavily on synchronized and centralized classical systems, making them susceptible to single-point-of-failure and synchronization-based attacks [23]. This centralized dependence undermines resilience and presents a significant vulnerability in the network architecture. Thus, developing robust security protocols and protective measures specifically for the quantum internet is crucial. Given the rapid advancements in quantum technologies and the emergence of new threats, these security measures must be continuously reviewed, revised, and adapted to ensure the quantum internet's long-term integrity and resilience.

To facilitate research in quantum internet security, this article provides an overview that not only compiles known vulnerabilities and mitigation strategies but also adds empirical value through simulations. It begins with a review of the core components of the quantum internet, followed by a layer-based structure that encapsulates the essential features of various quantum internet stacks [24]. We subsequently dissect vulnerabilities specific to each layer, offering initial detection, counteraction methods, and evaluation metrics governed by the key tenets of information security: Confidentiality, Integrity, and Availability (CIA). It is worth noting that this article does not attempt to offer a definitive solution. Instead, it highlights its significance through analytical findings and introduces a security framework for future research. Overall, the main contributions of this paper can be summarized as follows:

- We provide a layer-by-layer breakdown of the specific security issues encountered in quantum networks.
- We introduce various defensive strategies designed to address these issues across various layers.
- We assess the severity of potential attacks and evaluate the anticipated effectiveness of mitigation methods through analyses and simulations.
- We identify the vulnerabilities within different types of networks.

The rest of the paper is organized as follows. In the next section, we introduce the components and architectures of the quantum internet. Subsequently, we discuss the technical stack of the quantum internet. Afterward, we present a taxonomy of security concerns and initial countermeasures in the quantum internet. Finally, we discuss readiness of attacks and conclude.

## II. Elements of The Quantum Internet

This section lays the groundwork for analysis by providing a brief overview of the quantum internet's core components.

### A. Quantum Computers

Quantum computers operate on qubits, which fundamentally differ from classical bits. While classical bits are confined to discrete states of 0 or 1, a qubit can simultaneously exist in a superposition of both states, represented as $|\psi\rangle = \alpha|0\rangle + \beta|1\rangle$, where $\alpha$ and $\beta$ are complex numbers. The probabilities of measuring the qubit in each state ($|0\rangle$ or $|1\rangle$) are given by $|\alpha|^2$ and $|\beta|^2$, respectively, with $|\alpha|^2 + |\beta|^2 = 1$.

When multiple qubits interact under specific conditions, they can become entangled—a quantum phenomenon in which their states are strongly correlated, making it impossible to describe the state of any individual qubit independently of the others, regardless of the distance separating them. A notable example is the Bell states, representing a class of maximally entangled 2-qubit states [25], which can be represented as $|\phi_1\rangle = \frac{1}{\sqrt{2}}(|00\rangle \pm |11\rangle)$ or $|\phi_2\rangle = \frac{1}{\sqrt{2}}(|01\rangle \pm |10\rangle)$. In these states, a measurement performed on one qubit immediately determines the state of the other. For example, in the $|\phi_1\rangle$ state, if the first qubit is measured as $|0\rangle$, the second qubit will also collapse to $|0\rangle$. Conversely, if the first qubit is $|1\rangle$, the second will instantaneously become $|1\rangle$.

In addition to superposition and entanglement, interference is a crucial phenomenon that enables quantum algorithms to amplify the probabilities of correct solutions while minimizing those of incorrect ones. [26]. With interference, quantum states can be represented as wave functions with amplitudes that can interfere constructively (enhancing probability) or destructively (reducing probability). For instance, Shor's [27] and Grover's [28] algorithms use interference to refine the solution space, increasing the probability of achieving desired outcomes upon measurement. Shor's algorithm can factorize large integers exponentially faster than the best-known classical algorithms. It poses a significant threat to classical encryption systems, which rely on the difficulty of factoring large numbers. Meanwhile, Grover's algorithm provides a quadratic speedup for searching an unsorted database or solving unstructured search problems. For $N$ possible solutions, Grover's algorithm finds the correct one in $O(\sqrt{N})$ steps, compared to $O(N)$ for classical approaches. These algorithms exemplify the specialized computational capabilities of quantum computing, providing solutions to problems that are otherwise intractable for classical systems and emphasizing their transformative impact on fields such as cryptography, optimization, and pattern matching.

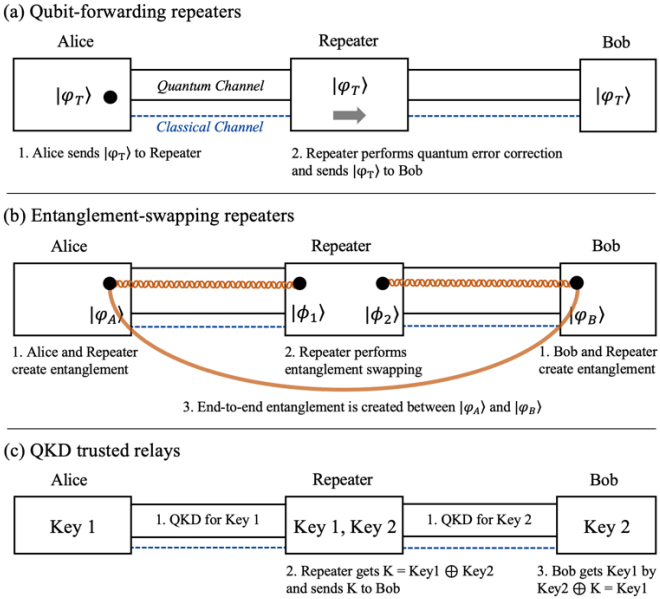

**Fig. 1.** Common types of quantum repeaters and key relays.

*B. Quantum Repeaters and Relays*

Qubits are highly sensitive to environmental interactions, making them prone to unintended alterations during storage and transmission. This susceptibility leads to decoherence and noise, often resulting in significant information loss. Quantum repeaters play a critical role in enabling long-distance quantum communication by mitigating these challenges. Unlike classical repeaters, which can freely amplify and retransmit signals, quantum repeaters are limited by the no-cloning theorem, which prohibits the duplication of quantum states. In this paper, three widely discussed repeater schemes are examined, as illustrated in Fig. 1: qubit-forwarding repeaters, entanglement-swapping repeaters, and QKD-trusted relays. Each method relies on a combination of authenticated classical communication and a quantum channel between nodes to function properly.

Qubit-forwarding repeaters function similarly to classical repeaters by forwarding data and correcting errors during transmission. They employ Quantum Error Correction (QEC) [29] to transmit encoded qubits in multi-photon states resistant to photon loss. As shown in Fig. 1a, encoded qubits undergo QEC at each repeater before being forwarded to the next node. This helps mitigate the impacts of noise and decoherence during transmission. However, an inherent challenge is that, even with QEC, noise can mask signs of eavesdropping, making detection more difficult. In other words, the end users might be unable to distinguish between noise-induced errors and interference from an eavesdropper. Nonetheless, qubit-forwarding remains a viable option for certain scenarios, particularly in networks with nodes separated by relatively short distances. This practicality is supported by the ability to adapt classical techniques and the ongoing advancements in QEC technologies.

Alternatively, entanglement-swapping repeaters facilitate the creation of end-to-end entanglement over long distances. This end-to-end entanglement serves as a communication resource for end users, allowing them to perform communication such as quantum teleportation or applications like QKD. The process involves several steps: initially, entanglement is generated between neighboring nodes using direct quantum links. The repeaters then perform entanglement swapping, an operation where two pairs of entangled qubits are combined to produce a single pair of qubits that are entangled across distant nodes without a direct physical connection. As shown in Fig. 1b, the repeater processes one qubit from each of two entangled pairs to merge into a single end-to-end pair that directly "connects" the two end users. This scheme has been demonstrated to be more effective for scaling quantum networks in the NISQ era than qubit-forwarding methods [30]. It does not require the large number of qubits needed for QEC as in qubit-forwarding. The requirements for purification (to achieve higher-quality entanglement) are significantly lower than those needed for error correction at each intermediate node [31].

Moreover, networks designed specifically for QKD often employ trusted relays to extend the range of key generation. These relays enable the transmission of QKD-generated keys between adjacent nodes using encryption. As shown in Fig. 1c, Alice generates a shared key (Key 1) with her neighboring repeater via QKD, while the repeater generates a second key (Key 2) with Bob via QKD. The repeater then encrypts Key 1 using Key 2 and transmits it to Bob over a classical channel. Bob subsequently decrypts the transmission using Key 2 to retrieve Key 1, which can be used for secure communication between Alice and Bob. While this approach allows for secure key exchange, it relies on the assumption that the repeaters themselves are trusted. Efforts are ongoing to eliminate trust dependencies and develop advanced protocols that address the rate-distance limitations of traditional QKD systems, such as twin-field QKD protocols [32]. It enables key distribution over long distance using a single-repeater architecture. It leverages the interference of coherent quantum states, offering a solution to extend the scalability and efficiency of QKD networks.

Each of these repeater schemes offers unique advantages and limitations. Qubit-forwarding repeaters prioritize QEC, which is more general but with certain performance constraints. Entanglement-swapping repeaters focus on entanglement-based applications, offering superior performance. QKD trusted relays, on the other hand, provide a practical solution for a secure key exchange network. Together, these approaches form the foundation for the quantum internet in the NISQ era.

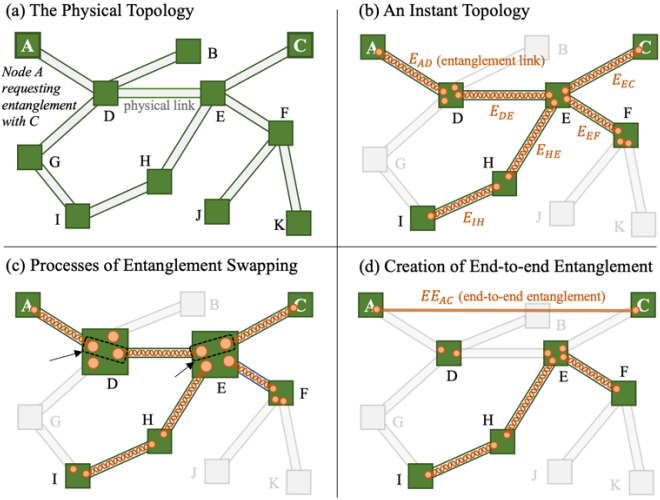

**Fig. 2.** Example of end-to-end entanglement distribution.

*C. Entanglement Routing*

As the number of network nodes grows and the topology extends beyond a linear chain, a routing scheme becomes essential to determine paths between arbitrary end nodes within the network. Classical routing techniques are largely applicable to qubit-forwarding and QKD relays. Entanglement-swapping repeaters, however, must account for additional constraints, including decoherence, unreliable quantum operation, and link state propagation. This subsection explores the routing aspects of entanglement-swapping schemes, i.e., entanglement routing.

Consider a network with the physical topology shown in Fig. 2a, where two endpoints (e.g., Nodes A and C) aim to establish end-to-end entanglement. In the beginning, neighboring nodes connected by physical links attempt to generate direct-link entanglement. For example, it is attempted between Nodes A and D, and between Nodes D and E. Some attempts succeed while others fail, and the nodes connected by successful direct-link entanglements form what is known as the instant topology, shown in Fig. 2b. That is, Nodes A, D, E, C, F, H, and I, along with their associated entanglements (e.g., $E_{AD}$, $E_{DE}$, $E_{EC}$ etc.), constitute the instant topology. For end-to-end entanglement between Nodes A and C, intermediary nodes D and E perform entanglement swapping using the qubits from the direct-link entanglements connecting them. For instance, as shown in Fig. 2c, Node D executes entanglement swapping. If successful, it results in the entanglement between Nodes A and E. Subsequently, Node E performs another swapping, and upon success, Nodes A and C achieve end-to-end entanglement, $EE_{AC}$, as shown in Fig. 2d. During this process, the entangled states $E_{AD}$, $E_{DE}$, and $E_{EC}$ are consumed. To serve subsequent end-user requests, neighboring nodes continuously attempt to regenerate direct-link entanglements.

The entanglement routing problem involves determining the optimal path(s) between end users to establish end-to-end entanglement(s). In the example provided in Fig. 2, the path A→D→E→C is the desired route, as it is the only feasible option. However, additional challenges arise in a more complex network where the instant topology is constantly changing due to state decoherence and consumption. Existing approaches suggest two types of methods: one utilizes synchronization to maintain a relatively stable instantaneous topology [14], [33], while the other employs asynchronous strategies to adaptively manage the topology in a distributed manner [15], [34]. Neither method effectively addresses the presence of malicious nodes, which severely compromise availability. Moreover, scenarios with concurrent user requests exacerbate these vulnerabilities, further complicating the routing process and heightening the risk of network disruption.

*D. Quantum Key Distribution*

Given the significant number of existing attacks targeting QKD, this subsection provides a brief introduction to how QKD functions. As discussed, QKD establishes classical secret keys between end users in quantum settings and offers the unique ability to detect intrusions through quantum measurements.

The BB84 protocol, introduced by Bennett and Brassard in 1984, is a cornerstone of QKD [5]. The underlying principle is that when a qubit is measured using the same method (referred to as a measurement basis), it consistently yields the same result. However, if measured on a different basis, the outcome might vary. Let us see a simplified example of BB84, where Alice wants to share a secret key with Bob:

1. Preparation: Alice generates a random bit sequence and translates them into qubits using either rectilinear (i.e., $|0\rangle$ or $|1\rangle$) or diagonal (i.e., $|+\rangle$ or $|-\rangle$) bases, and sends them to Bob.
2. Transmission: Alice sends her qubits to Bob via a quantum channel.
3. Measurement: Bob randomly selects a basis for each incoming qubit and records the result.
4. Basis Comparison: They share their chosen bases (not results) publicly. For matching bases, the qubit values should align.
5. Key Generation: Mismatched basis measurements are discarded, leaving a shared secret key.

BB84 and its variants are commonly referred to as "prepare-and-measure" protocols. In these protocols, a portion of the key is shared publicly to detect eavesdropping. Any discrepancies in the disclosed key indicate tampering, leading the end users to disregard the compromised channels. Prepare-and-measure protocols remain vulnerable to DoS attacks.

Following the BB84 protocol, various QKD protocols have been developed to address vulnerabilities and enhance security. In particular, the E91 protocol [37], proposed by Artur Ekert in 1991, and its variants are known as entanglement-based QKD protocols. In this approach, a third party generates pairs of entangled states and distributes them to two end users. Similarly, by measuring their respective qubits and comparing the publicly shared measurement bases, they keep the outcomes with matching bases (where both used the same basis) and discard the rest. The entanglement ensures that any eavesdropping attempt disrupts the measurement outcomes, thereby exposing its presence. Nonetheless, vulnerabilities still appear across various layers of the quantum internet. More examples include recent protocols such as device-independent

5QKD (DI-QKD) [35], and measurement-device-independent QKD (MDI-QKD) [36]. These protocols are designed to mitigate potential attacks that exploit the weaknesses in prepare-and-measure schemes like BB84, offering stronger security guarantees in practical implementations.

## III. THE QUANTUM INTERNET STACK

Like its classical counterpart, the quantum internet can be organized into distinct layers. We define a simplified layer-based structure based on the core features of various quantum internet stacks [24]: physical, link, network, and application. Rooted in recent advancements, these layers offer specific services and protocols, as detailed in Table I. Note that each layer also incorporates components of classical networks.

At the bottom, the physical layer forms the foundation of the quantum internet, consisting of storage, computation, and transmission devices. This layer generally includes quantum memories, computing units, and communication links. While there are different types of repeaters, quantum networks often employ systems based on photons, using optical fibers or open-air transmissions [9].

Above the physical layer, the link layer manages direct connections between neighboring nodes, i.e., handling qubit transmission, or generating direct-link entanglement between them. In this layer, qubit-forwarding repeaters manage the physical transfer of data and QEC between adjacent nodes. Entanglement-swapping repeaters create entanglement between non-adjacent nodes [1]. Additionally, this layer facilitates remote QKD processes in settings involving trusted relays. Overall, the link layer is crucial for maintaining the integrity and security of data as it moves from one node to another, ensuring that quantum information is accurately relayed, or that entanglement is preserved during transmission.

The network layer manages the routing and flow of quantum traffic throughout the network, which varies based on the types of repeaters used and may involve qubit transmissions or entanglement requests. Although applying traditional networking methods to qubit transmissions may appear straightforward, the process often requires many QEC qubits. This has made the entanglement-swapping routing method a key area of research focus. Nevertheless, the practical implementation of end-to-end entanglement routing is fraught with challenges, including limited coherence times and the complexities involved in unreliable operations. These issues have spurred further research into quantum routing protocols and resource management [14], [15], [38].

At the top of the stack, the application layer provides users with quantum applications and services, including quantum teleportation, sensing, and distributed quantum computing. It is important to note that the placement of QKD within the network architecture can vary depending on the technologies used: in networks based on QKD relays, it is positioned at both the link and application layers, whereas in entanglement networks, it is considered part of the application layer.

TABLE I
PROTOCOL STACK OF QUANTUM INTERNET

| Layers | Components | Example Technologies |
|---|---|---|
| Application Layer | Quantum cryptography | End-to-end QKD. |
| | Data transmission | Quantum teleportation. |
| | Large-scale computing | Distributed computing. |
| Network Layer | Quantum router | Routing protocols. |
| Link Layer | Quantum repeater scheme (direct-link protocol types) | Qubit forwarding. Entanglement swapping. QKD relay. |
| Physical Layer | Transmission medium (quantum channel and classical channel) | Optical fiber. Free space/Satellite. |
| | Quantum computer | Photon-based. Superconductor-based. |

## IV. VULNERABILITIES AND COUNTERMEASURES

In alignment with the quantum internet stack discussed above, we categorize quantum internet security risks into four main areas: physical, link, network, and application layer attacks. Each category contains attacks specific to its respective layer.

### A. Physical Layer

Physical layer attacks often focus on quantum memories, communication resources, and other hardware components. These vulnerabilities typically stem from flaws or limitations in the hardware. In this section, we discuss attacks that target the physical layer.

*A.1. Photon Number Splitting (PNS) Attacks*

PNS attacks are frequently associated with the "prepare and measure" QKD protocols, such as BB84, which requires transmitting quantum states. These attacks, however, can target any photon-based qubit transmission, either wired or free space.

In PNS attacks, an eavesdropper capitalizes on the fact that a sender might occasionally emit multiple photons when using a weak coherent pulse source [20], [39]. This means some pulses may contain no photons, some a single photon, and some multiple photons. In the case of multiple photons, an eavesdropper could intercept some of them, measure them to gain information about the transmitted state, and then pass the remaining photons to the receiver undetected, thus compromising confidentiality.

PNS attacks can be countered through various methods [40], [41], [42]. One direct approach is to use a true single-photon source rather than an attenuated laser. However, this technology is still evolving [43]. While a single-photon source has been successfully experimented with [44], it faces challenges due to relatively low efficiency. This inefficiency arises from the quantum nature of the emission process, where the probability of producing a photon at a desired time is inherently limited. This results in a lower rate of photon production compared to more conventional sources such as attenuated lasers.

#

An alternate strategy is decoy states [42], which use states (pulses) with varying intensities: signal states and one or more decoy states. The signal states are used for the key distribution, while the decoy states detect eavesdropping. The key idea is that the sender randomly varies the intensity of the laser pulses between signal and decoy states without informing the receiver about which pulses are which. The decoy states are typically sent at different, often lower, intensities than the signal states. The lower intensity of the decoy pulses means they are less likely to contain multi-photon instances, making them ideal for detecting eavesdropping. If an eavesdropper tries to intercept these pulses, legitimate users can detect the differences in the quantum statistics between the intercepted decoy and untouched signal pulses. By comparing the rates of quantum bit errors and other parameters between signal and decoy pulses, legitimate users can estimate the fraction of single-photon pulses and detect the presence of an eavesdropper. This analysis allows them to adjust their key generation processes to account for any potential information leakage and ensure the security of the final key.

Furthermore, employing MDI-QKD or entanglement-based QKD can also prevent such attacks. In MDI-QKD, Alice and Bob prepare an entangled pair and send one half to a third party, Charlie, who is not necessarily trusted. Charlie performs a joint quantum measurement on the incoming qubits from Alice and Bob, similar to entanglement swapping. He then publicly announces the outcome of his measurement, which doesn't disclose any specific information about the states sent by Alice and Bob. Rather, it only confirms whether the qubits from Alice and Bob are correlated in a specific way as per the Bell state measurement. Based on Charlie's announcement, Alice and Bob can identify which of their qubits were successfully entangled during the measurement and disregard any unentangled qubits. By leveraging the principles of quantum entanglement and nonlocality, and not depending on the reliability of the measurement devices, MDI-QKD and entanglement-based QKD are considered secure against PNS and other detector side-channel attacks.

Finally, to assess how a PNS eavesdropper's actions could influence their detection, we conducted a simulation analyzing the distribution of photons Bob receives under PNS attacks. In the simulation, Alice sends Bob photons distributed according to a Poisson distribution with an average value of 5, which mimics the random photon counts typical of weak coherent pulses. Eve, the eavesdropper, is assumed to be able to intercept some photons and forward the remainder. We explored three different scenarios, depicted in Fig. 3:

1. Without Eve: Bob receives all photons sent by Alice.
2. With Eve (Random): Eve randomly intercepts photons, and the rest goes to Bob.
3. With Eve (Always -1 Photon): Eve takes one photon, forwarding the rest to Bob.

Fig. 3 illustrates how Eve's interference alters the photon distribution received by Bob. The Z-score graph within Fig. 3 highlights the differences between the photon distributions, which indicates Eve's presence. A higher Z-score indicates a more substantial difference. Eve can hide better in the "Always -1 photon" setting compared to the "Random" setting (when only considering distribution differences), although the latter may provide more information. Bob, however, requires multiple transmissions to confirm this altered distribution.

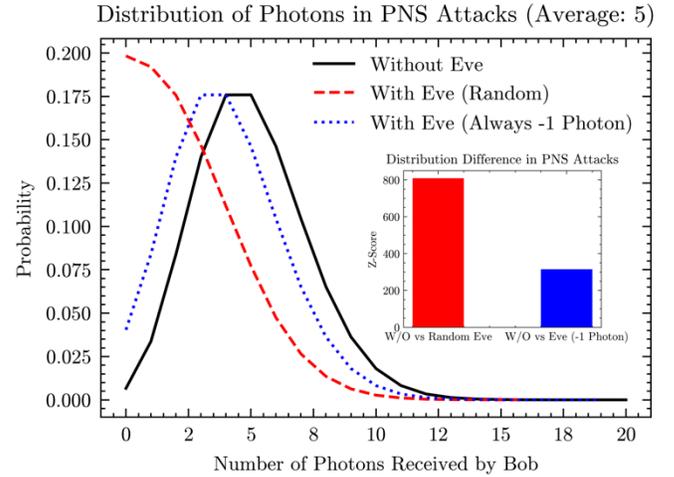

**Fig. 3.** Photon distribution in the simulation of PNS attacks.

*A.2. Trojan-Horse Attacks*

In the context of quantum networks, a Trojan-horse attack involves an attacker sending photons that appear harmless to the recipient but are embedded with harmful signals or configured in a manner that allows the attacker to extract information or disrupt the system when these photons are measured or processed by the recipient's equipment.

With unauthorized access to a quantum link, an attacker can initiate Trojan-horse attacks by sending a bright light pulse through the channel and examining the reflected pulses to gather information on the transmitted quantum states [45], [46], [47], [48]. This is possible because no optical component perfectly transmits or absorbs light, allowing some light to travel in the opposite direction of the main optical signal. This vulnerability enables such attacks, for example, to determine the measurement basis used in BB84 or acquiring information about transmitted qubits. These breaches can compromise both the confidentiality and availability of communications.

Taking "prepare and measure" QKD as an example, to counter such attacks, Alice and Bob need to discard a substantial portion of their raw key through a process known as privacy amplification [20]. It involves distilling a shorter, highly secure key from the longer, potentially compromised raw key using specific mathematical functions (e.g., hash functions) to eliminate Eve's partial knowledge. This may, however, reduce the secret key rate.

To reduce Eve's photon gain without significantly lowering the key rate, Alice and Bob can employ phase randomization on the transmitted photons, introducing a layer of randomness and uncertainty to counter these attacks [49]. The phase of a quantum state is a critical characteristic that defines the state, along with the probability amplitude. While changing the phase does not affect the probabilities of measurements in the $|0\rangle$ and $|1\rangle$ basis, it influences the outcomes in other bases used to describe the state. By randomizing the phase, the quantum state undergoes unpredictable shifts. The altered phase would mask its true state if Eve intercepted the qubit. Alice and Bob can then synchronize their phase-randomized states over classical channels.

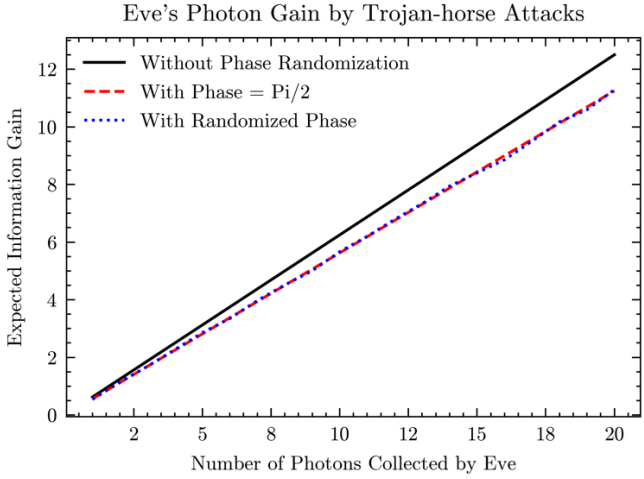

**Fig. 4.** Phase randomization in the Trojan-Horse simulation.

Eve must guess the measurement basis when she intercepts a reflected photon to extract information through these attacks. For clarity, let us define Eve's information gain as the number of photons she successfully measures on the correct basis plus the partial information she obtains when measuring in an incorrect basis, denoted as $G$. That is, a correct guess provides a complete photon gain: $G + 1$. An incorrect guess yields a partial photon gain: $G + \gamma$ where $0 \leq \gamma < 1$.

In Fig. 4, we plot Eve's photon gain as a function of the number of photons she collects during transmissions from Alice to Bob, comparing scenarios with and without phase randomization. In the simulation, Alice can send four possible states: $|0\rangle$, $|1\rangle$, $|+\rangle = \frac{1}{\sqrt{2}}(|0\rangle + |1\rangle)$ and $|-\rangle = \frac{1}{\sqrt{2}}(|0\rangle - |1\rangle)$. The states $|0\rangle$ and $|1\rangle$ are on a rectangular basis, while $|+\rangle$ and $|-\rangle$ are on a diagonal basis. As discussed, the photon gain can be calculated as follows:

$$G = \sum_{i=0}^{n} \gamma_i \quad (1)$$

where $n$ is the number of photons collected by Eve and $\gamma_i$ represents the incremental gain from measuring the $i$th photon, with $0 \leq \gamma_i \leq 1$. $\gamma_i$ depends on Eve's probability of correctly guessing the measurement basis used by Alice and is given by:

$$\gamma_i = \begin{cases} 0.5, & b_i^E \neq b_i^A \\ 1, & b_i^E = b_i^A = 1 \\ \frac{1}{2}\cos^2\frac{\theta_i}{2} + \frac{1}{2}\sin^2\frac{\theta_i}{2}, & b_i^E = b_i^A = -1 \end{cases} \quad (2)$$

where $b_i^E$ and $b_i^A$ represent the measurement bases used by Eve, and Alice for the $i$th photon, respectively, with values from the set $\{1, -1\}$ where 1 is the rectangular basis, and -1 is the diagonal basis. $\theta_i$ is the phase shift Alice applies to the $i$th photon.

As mentioned, phase randomization does not influence the photon gain when using the rectangular basis but reduces the gain in the diagonal basis. Without phase randomization, Eve can precisely determine the state of a photon intercepted from the diagonal basis ($|+\rangle$ or $|-\rangle$). With phase randomization, each photon Alice sends on this basis receives a random phase shift, rendering states $|+\rangle$ and $|-\rangle$ unpredictable (i.e., lowering Eve's gain). If Eve intercepts and measures such a photon, her chances of accurately identifying the state drop to $\frac{1}{2}\cos^2\frac{\theta_i}{2} + \frac{1}{2}\sin^2\frac{\theta_i}{2}$ because, depending on whether the state is $|+\rangle$ or $|-\rangle$, Eve's probability of being correct reduces to $\cos^2\frac{\theta_i}{2}$ or $\sin^2\frac{\theta_i}{2}$, respectively.

With that, this simulation compares scenarios without phase shifts, with a consistent $\pi/2$ phase shift, and with randomized phase shifts for each photon. Notably, if Eve intercepts a photon with a $\pi/2$ phase shift on a diagonal basis, her chance of a correct measurement is 50%. The results highlight the effectiveness of phase randomization in protecting against Trojan-horse attacks, with overlapping performance between randomized phase shifts and consistent $\pi/2$ phase shifts.

*A.3. Other Attacks on Quantum Devices*

In addition to the two well-known attacks discussed above, there are other side-channel strategies targeting the imperfection in current quantum devices, including interference or tampering with the physical components utilized in communications, such as light sources, detectors, or quantum storage, to risks. For instance, photon-based systems heavily depend on detectors to measure the quantum properties of single photons. Yet, these detectors are susceptible to compromise through specifically tailored bright illumination [50], [51]. Experiments showed that Eve could manipulate the gated avalanche photodiodes in the systems using bright illumination, effectively blinding them and transforming them into classical, linear detectors [50]. This alteration allows Eve to control the detector outputs, potentially capturing the full secret key undetected. Countermeasures such as installing optical power meters at detector inputs and incorporating their readings into security proofs can help thwart such blinding attacks. Nonetheless, determining completely hack-proof configurations for detectors remains a challenging task.

Additionally, phase-remapping attacks involve an attacker manipulating the phase settings within a quantum communication channel. This manipulation changes the intended phase encoding of the qubits being transmitted between legitimate parties, potentially leading to incorrect decoding and creating vulnerabilities in communication. The attack was demonstrated through experiments using a standard commercial QKD system that relies on phase-encoded photons [52]. The phase modulators in the system were altered to show how an external attacker could affect the phase settings during the key distribution process. By adjusting the voltages applied to these modulators, the phases of the encoded qubits were changed without being detected by the standard security checks in QKD. This manipulation could cause the receiver to misinterpret the quantum states, leading to errors in the key extraction process.

Moreover, strategies against such attacks include using quantum tamper-evident seals [53]. The goal of quantum seals is to leverage quantum mechanics to detect and indicate any unauthorized access or tampering. However, quantum states are highly sensitive to environmental disturbances such as temperature, electromagnetic fields, and movements. While useful for detecting tampering, this sensitivity also makes the seal prone to false positives from non-malicious environmental changes. Additionally, some setups incorporate multiple quantum systems as backup channels or systems to increase availability in case of system compromise or failure. If one system experiences a failure or a security breach, the network can automatically switch to an alternative system, thus ensuring continuous secure communication. For instance, the QKD network in Vienna utilizes various QKD systems, each based on different principles of quantum mechanics, such as measurement or entanglement [54].

*B. Link Layer*

Attacks at the link layer target the protocol that governs communications or direct-link entanglements. This underscores the necessity for robust link layer protocols to protect against unauthorized state interference during transmission. This protection includes secure state encoding and error detection to rectify transmission anomalies.

*B.1. Entangling-probe Attacks*

In an entangling-probe attack, the attacker replaces the initialized qubits in a user's quantum register with qubits entangled with those in the attacker's register. This manipulation allows the attacker to perform local operations on these qubits and potentially alter the output of the user's algorithm [55], [56]. For instance, after the replacement, the attacker establishes entanglement between the qubits in the user's register and the qubits they control. Suppose the user executes an algorithm like Shor's algorithm with their qubits. In that case, the attacker can interfere with the operations, leading the user to obtain a random result by manipulating the entangled qubit, thereby compromising the integrity and availability. Also, the attacker might remain passive and wait for the user to measure their qubit. Once the measurement occurs, the attacker, through their entangled qubit, can deduce the user's results, thus breaching confidentiality.

In another scenario, suppose a user, Bob, already maintains an entanglement with another user, Alice. As illustrated in Fig. 5, if an external entity, Eve, gains access to Bob's quantum computer, she can entangle her qubit with the pair. This can be done by applying a CNOT gate to Bob's qubit and her own, as depicted in the circuit in Fig. 5. Then, the qubits of Alice, Bob, and Eve become entangled in a tripartite state. Any interaction by one party affects the entire entangled state and influences the other parties involved. If Alice and Bob believe they are using a secure direct-link entanglement, Eve could interfere with or eavesdrop on their communications. For instance, if Alice and Bob use their entanglement to generate a secret key, Eve might also be able to access it.

Additionally, there are potential methods for eavesdropping during the generation and delivery of entangled photon pairs. For instance, in typical direct-link entanglement generation scenarios, entangled photons are produced by a third party and then sent to Alice and Bob. If present, Eve could attack using either a polarizer or a decohering birefringent plate, both of which can be oriented and, in some cases, accompanied by additional wave plates to allow analysis in arbitrary elliptical polarization bases. Experiments have demonstrated that the presence of an eavesdropper can be consistently detected by checking for violations of Bell's inequalities [57].

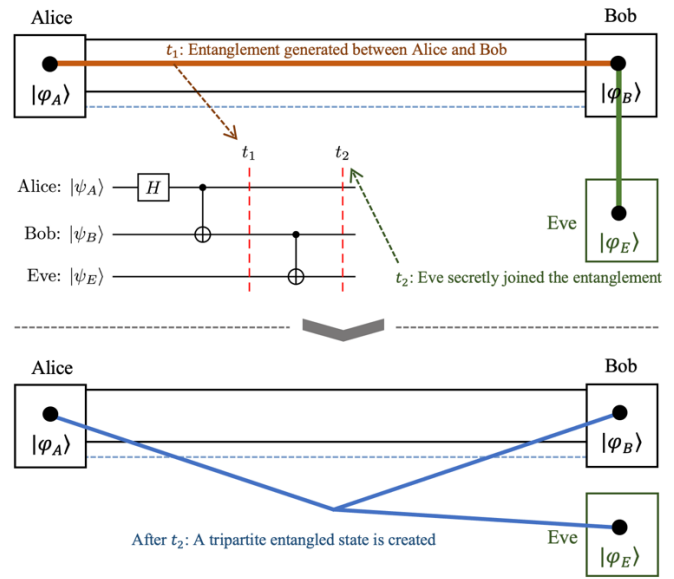

**Fig. 5.** An infiltration with entangling probe.

*B.2. Man-in-the-middle Attacks*

In addition to entangling probe attacks, if an attacker gains access to a user's computer, they can launch MiTM attacks by impersonating the receiver or sender. For example, Eve can deceive Alice into believing she is communicating with Bob when she interacts with Eve. This deception can be accomplished by intercepting both classical and quantum channels. Eve can then manipulate the intercepted classical messages to align with her goals—copying, modifying them, or creating entirely new messages unrelated to Alice's original. She can also replicate the initial protocol used to re-encode the message intended for Bob in the next communication phase and send it to him.

In classical networks, authentication is a standard defense against MiTM attacks. However, authenticating in quantum systems presents challenges. Alternatively, Alice and Bob could use an interlock protocol [58], a method commonly discussed in classical networks for detecting the presence of Eve. In this protocol, Alice encrypts her message with a code that cannot be decoded instantly, and Bob does the same. Alice sends a portion of her encrypted message to Bob, and he sends a portion of his to Alice. Upon receiving Bob's partial message, Alice sends the remaining part of her encrypted message; similarly, upon receiving Alice's part, Bob sends the remainder of his. The messages are then combined to complete the communication. This sequential exchange prevents Eve from altering any part of the message without having access to the entire message. If Eve attempts to modify and send a previous part to Bob, she must accurately predict the subsequent part that

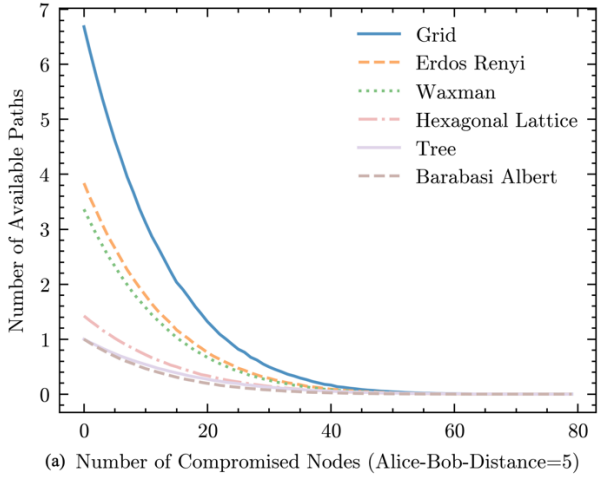

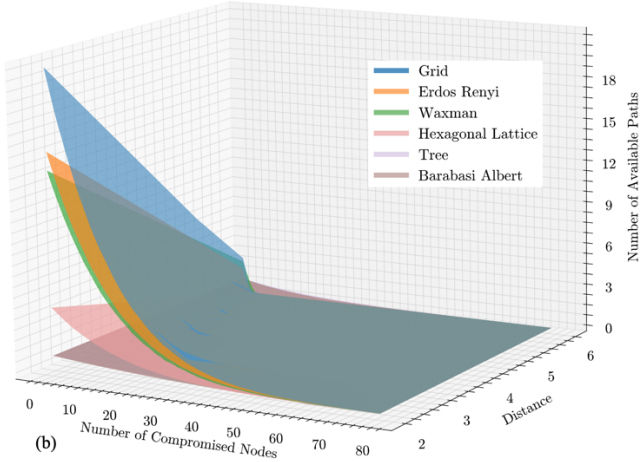

**Fig. 6.** 100-node network simulation results across various topologies (topologies detailed in [62]).
*Parameters: Gird ($10 \times 10$), Erdős Rényi (probability for edge creation, $p = 0.2$), Waxman (model parameters, $\alpha = 0.1$, $\beta = 0.4$), Hexagonal Lattice ($6 \times 7$), Tree (height $h = 4$, children of each node $r = 3$), Barabási Albert (number of neighbors, $k = 3$). (a) Decrease in path numbers as untrusted repeaters grow; (b) 3D plot by adding the distance dimension.*

Bob expects. An incorrect prediction will result in a nonsensical reconstructed message, alerting Bob to potential tampering. While in classical networks, this method does not prevent Eve from merely listening and forwarding, in quantum networks, any attempt to read or copy the quantum state would collapse, revealing the interference [59].

*B.3. Attacks on Error Correction*

Qubit-forwarding transmission relies heavily on QEC, which introduces vulnerabilities. If adversaries know the specific QEC scheme, they can introduce noise to disrupt it. They might deliberately introduce errors just below the QEC code's tolerance threshold, potentially overloading its capacity and leading to logical errors. This critical point occurs when the rate of logical errors (errors affecting logical qubits) exceeds the rate of physical errors (errors affecting physical qubits).

One fundamental weakness is the susceptibility to specific types of errors [60]. For example, the 3-qubit bit-flip code, effective against single bit-flip errors, cannot handle multiple simultaneous errors. Also, it is ineffective against phase-flip errors that transform states, such as $|+\rangle$ to $|-\rangle$. Another vulnerability is that QEC's reliability hinges on maintaining an error rate below a break-even point. For the 3-qubit bit-flip code, this is 1/3. Breaching this point results in ineffective error correction. Also, QEC relies on periodic checks using ancillary qubits. Any delay or failure in these periodic checks can go unnoticed, leading to an accumulation of errors. QEC is also computationally intensive; an adversary could exploit this by forcing the system into near-constant error correction, essentially launching a DoS attack.

To address these vulnerabilities, a combination of QEC codes can be employed and alternated, complicating the ability of attackers to introduce noise [61]. Using redundant ancilla qubits, cross-referencing outcomes, and maintaining a secure environment are crucial strategies. Nonetheless, given the current hardware limitations associated with QEC in qubit transmissions, entanglement-swapping networks represent a more viable option in the near term.

*C. Network Layer*

At the network layer, attacks are aimed at disrupting operations and resources, including routing and entanglement generation. Routing strategies that depend on centralized control, for instance, are prone to single points of failure. This section details the vulnerabilities in quantum networking.

*C.1. Untrusted Repeater Nodes*

Most existing quantum routing schemes presume the use of trusted repeaters [14], [63], [13], [64]. However, this assumption introduces several vulnerabilities, particularly if an attacker gains access to a repeater or substitutes it with a counterfeit device. For instance, an attacker could deploy a fake quantum repeater or hijack an existing one to reroute traffic through a compromised system, akin to spoofing attacks in classical networks [65].

In qubit-forwarding networks, a breach of the quantum repeater would enable an attacker to intercept or manipulate all qubits passing through a compromised repeater. Once in control, the attacker could reroute network traffic, harvesting communication resources and injecting malicious payloads. Furthermore, diverting traffic through a malicious repeater could lead to network inefficiencies or enable targeted denial of service attacks. Detection of such tampering in these networks is relatively straightforward, as any interference with the transmitted qubits would cause the qubits to collapse, which the receiver would notice. However, effectively preventing such attacks is challenging due to the scarcity of practical encryption methods for qubits [66]. Presently, efforts are focused more on transmitting a reliable and larger number of qubits. This vulnerability can be mitigated by utilizing entanglement-swapping repeaters, which do not transmit data but establish end-to-end entanglement, thus circumventing potential MiTM attacks once the entanglement is set up.

Nonetheless, in networks that use entanglement-swapping, an attacker could potentially disrupt the entanglement process within quantum repeaters. By interfering with the entanglement process, an attacker could introduce errors before swapping, thereby reducing the success rate of the swaps and consequently lowering the overall rate of end-to-end entanglement [67]. This could also allow an attacker to modify its operations, such as altering how routing is processed, which could sabotage or manipulate the creation of end-to-end entanglement. Moreover, a compromised repeater might secretly introduce a third qubit into a two-party end-to-end entanglement and forward this extra qubit to an eavesdropper, as discussed in Section IV.B.1. This could allow unauthorized parties to listen in on or interfere with the operations performed on the end-to-end entanglement.

Additionally, for the routing and swapping process to be complete, there needs to be an entanglement identifier and quantum repeater identifiers [68], [69], [70]. However, the inherent complexities of quantum systems, such as their sensitivity to disturbances and the presence of quantum noise, make it difficult to apply traditional control and identification methods. The lack of robust identification mechanisms for quantum systems exacerbates the problem [71], [72]. An attacker could impersonate a quantum repeater by forging identifications and undertaking malicious activities such as selecting a longer or inaccessible path. They might send fraudulent addressing messages to quantum repeaters, linking a malicious address with that of a legitimate quantum repeater in the network. This could result in operations intended for the legitimate address being redirected and tampered with by the attacker.

Regardless, when a repeater is compromised and identified, a common solution is to bypass it, although this may introduce additional hops. We conducted simulations using various network topologies to assess the impact of bypassing untrusted repeaters on performance. We created networks with approximately 100 nodes in different configurations, including a $10 \times 10$ grid and a four-level balanced tree where each node has three children. We also generated random topologies using models such as Erdős-Rényi, Waxman, and Barabási-Albert [62]. In each network, we randomly designate certain nodes as compromised and exclude these nodes when finding viable paths. As expected and shown in Fig. 6, the simulations indicated an exponential decrease in available paths as the number of compromised repeaters increased. When the proportion of untrusted nodes surpasses a certain threshold—such as 60% of the network being compromised—all viable paths become unavailable, as shown in Fig. 6b. Moreover, some network topologies demonstrate more resilience to compromised nodes than others. For example, grid topologies, with their non-hierarchical structure and multiple connection routes, are inherently more robust against node failures or compromises than tree topologies.

The analysis assumes that compromised nodes have been identified; however, assessing a node's trustworthiness remains challenging. Thus, a set of Quality-of-Service (QoS) metrics for quantum networks is needed to identify abnormal network patterns that may indicate a compromised node [73], [64], [69]. For example, nodes with higher error rates and longer response times would be considered untrustworthy. These QoS metrics could include critical performance indicators like throughput, delay, and overall utilization of network resources, which may vary depending on the type of network. For instance, throughput could be measured by the number of qubits transmitted in qubit-forwarding networks. In contrast, in entanglement-swapping networks, it could be determined by the rate of end-to-end entanglement generation. Based on these criteria, network paths could prioritize nodes that meet specific reliability standards and put weights in pathfinding algorithms, similar to the common practices in classical networks.

Finally, it is worth noting that, in MDI-QKD setups, relays can be untrusted because the protocol addresses vulnerabilities related to detector side channels. The role of the relay is limited to announcing the results of the measurements rather than the measurements themselves, preventing potential eavesdroppers, including those with malicious intent operating the relay, from accessing sensitive information. However, these untrusted relays function merely as single repeater links connecting two adjacent nodes. Therefore, unlike trusted relays, MDI-QKD does not support QKD over arbitrarily long distances [16]. For large-scale QKD networks, integrating both trusted and untrusted relays is expected to be essential.

*C.2. Denial of Service*

DoS attacks can target multiple layers within a quantum network. At the physical layer, such attacks may involve interference with the quantum communication infrastructure, such as intercepting or altering qubits. While the end parties can detect an attacker's presence, their only solution is to switch to a different channel, thereby disrupting the service of the original communication channel. At the network layer, DoS attacks could interfere with routing quantum information or flood the network with excessive connection requests, leading to degraded performance. In this subsection, we focus on DoS attacks in the network layer.

In connection-oriented quantum networks [38], [74], a queue of pending connection requests is managed by a centralized routing system. The system processes these requests either individually or in batches. An attacker can continually submit connection requests to exhaust network resources or inundate its processing capability. Although connectionless networks might mitigate this issue by allowing a request to avoid blocking other connection requests through a dedicated connection path, they are not entirely immune and still face similar threats when congestion occurs [74].

Filtering suspicious traffic that causes DoS at the physical layer is challenging in quantum networks. Although the quantum measurement property allows us to detect the presence of an eavesdropper, there is no straightforward method to eliminate them. Thus, the only strategy is to utilize backup channels for availability. However, DoS attacks can be countered at the network layer using traditional defense mechanisms, as these attacks generally originate from classical networks within hybrid quantum-classical systems. For instance, DoS attacks involving excessive connection requests can be managed by scrutinizing these requests for abnormal patterns that suggest an attack. Once detected, the origins of these dubious requests can be isolated or redirected to a "sinkhole," where they are discarded. This approach mirrors the sinkhole technique used in classical networks [75].

Another widely used tactic involves rate limiting and implementing timeouts, which restrict the number of connection requests a single source can make within a specific timeframe. This method helps prevent both intentional and accidental DoS attacks originating from within the network by capping the number of incomplete or embryonic connections—a type of connection that has yet to complete the necessary classical communications between the source and destination, such as messages that communicate the results of direct-link entanglement generation [14], [63], or those detailing the instant network topology for routing purposes [15]. When the threshold for these embryonic connections is exceeded, further requests from the source are blocked until the network can verify or permit them again [76].

Integrating queuing or scheduling algorithms can improve the management of connection requests, mitigating congestion that might escalate to DoS. Adopting fair and efficient resource management strategies tailored to quantum QoS metrics is thus crucial [38]. For instance, schedulers could prioritize requests based on levels of suspicion to keep services running for legitimate users while minimizing the impact of an attack. This could involve assessing factors such as the frequency and type of requests a session generates and comparing these to the normal behavior patterns of legitimate users [77].

Similar to the solution to DoS at the physical layer, allocating backup resources for essential operations also helps the network sustain availability. By having multiple network paths and routing servers, even if one route is compromised or overwhelmed by a DoS attack, other paths can handle the necessary connections. This redundancy helps balance the load and provides alternative routes for data to travel, reducing the potential for a single point of failure [78], but it comes at a higher cost.

*C.3. Routing Disruption*

Most quantum routing schemes, such as those used in qubit-forwarding and entanglement-swapping networks, rely on a classical centralized routing node to perform tasks such as pathfinding and managing quantum memories across lossy links [38], [33]. These routing nodes are vulnerable to attacks, particularly when they are overloaded with an excessive number of requests. Such attacks could fill the routing table and exhaust the nodes' processing and memory resources, increasing response times and potentially leading to DoS. The reconnection attempts of failed requests could exacerbate this situation. If routing nodes are hijacked or impersonated, attackers could manipulate the routing process to divert qubits through compromised nodes they control, potentially leading to eavesdropping or data alteration and thus compromising confidentiality and integrity. Attackers might also force data to take longer paths, increasing loss rates and hindering successful communication. They could even introduce "shadow nodes" that appear legitimate, drawing users to route data through them, similar to how rogue Wi-Fi hotspots attract devices with their strong signals [79].

For entanglement distribution networks, security concerns vary based on whether there is global or local link-state knowledge [14]. With global link-state knowledge, every node has a holistic view of the direct-link generation status before routing (i.e., a global view of the instant topology in Fig. 2b), which typically requires a centralized routing server and shares the vulnerabilities mentioned above. With local link-state knowledge, nodes only have information about the direct-link entanglements with their adjacent nodes. Each node then indiscriminately swaps entanglements to establish an end-to-end entanglement between endpoints [80]. If an attacker hijacks a node, they can manipulate swaps to reduce the success rate, increasing delays and compromising availability. In distributed graph structures, such as those used in asynchronous routing schemes, certain nodes, like the root nodes of trees, may control the routing graph [15]. An attacker masquerading as a trusted root node could announce invalid routes, directing nearby user requests towards these non-functional routes and violating availability and integrity. This impact could be more severe if the attacker can hijack multiple routes across the network.

To combat these threats and ensure the integrity and availability of the network, it is crucial to implement redundant resources and robust detection mechanisms within classical networks, particularly when dealing with untrusted repeater nodes. For instance, employing a distributed routing system rather than relying on a centralized network node can avoid single point of failure [81], [82]. Additional protective measures like encrypting classical network traffic and implementing strong authentication and authorization are also vital. Additionally, attackers can target resource management algorithms that allocate quantum memories, attempting to monopolize these resources by mimicking criteria prioritizing connection requests. Reliable criteria and redundant resources discussed can help thwart these maneuvers.

Finally, an attacker could exploit vulnerabilities in time synchronization within quantum systems [83], altering the expected arrival times of entangled states and affecting entanglement-swapping outcomes. Many entanglement routing schemes depend heavily on event timing precision [14], [33]. Disrupting time synchronization could lead to discrepancies in the anticipated timing of entanglement generation, introducing errors and misinterpretations. One immediate solution is to eliminate synchronization in routing and adopt asynchronous routing protocols, such as [15], [80], [82]. Additionally, utilizing multiple independent time sources can diminish reliance on a single synchronization point. Each node can choose the most reliable time signal available, or a consensus mechanism could be employed to establish the correct timing by aggregating inputs from various sources.

*D. Application Layer*

At the application layer, attacks mostly target vulnerabilities in quantum applications such as QKD, quantum algorithms, or distributed quantum computing architectures. These attacks can take various forms, including tampering with the integrity of QKD, manipulating inputs or outputs of quantum algorithms, or disrupting the coordination of distributed quantum resources.

12#

## D.1. Quantum Algorithmic Attacks

Quantum applications often involve collaborative execution of algorithms or protocols, either by receiving information from one end or using end-to-end entanglement jointly [7], [84]. For instance, two distant nodes participating in a distributed quantum computing setup might need to execute a joint computation. These computations often rely on entanglement to perform operations across the network. One example is executing a distributed version of Grover's search algorithm, where nodes share database segments and use entanglement to enhance search efficiency across the distributed system. In another scenario, two remote entities, such as a bank and its branch office, might employ QKD to establish a secure channel. Additionally, two laboratories might collaborate to teleport a quantum state from one to another, necessitating a pre-shared entangled pair. In this process, one lab prepares a quantum state and performs a Bell state measurement using its segment of the entangled pair. The measurement results are then communicated classically to the other lab, which applies a specific transformation to their pair segment, effectively reconstructing the original quantum state. Such collaborative architectures introduce vulnerabilities similar to those encountered in classical distributed systems, including compromised nodes, challenges with consensus, and the risk of fault propagation [85].

Moreover, attacks on quantum applications often target protocols, algorithms, and architecture vulnerabilities. For instance, amplitude amplification in Grover's algorithm enhances the probability of desired states within a superposition, increasing their likelihood of detection and providing computational benefits [28]. However, while essential for the algorithm's effectiveness, this amplification can also inadvertently magnify any quantum states or operations errors. These errors might stem from imperfect quantum gates, environmental decoherence, or errors in initial state preparations, potentially concealing an attacker's operations. An attacker could exploit such vulnerabilities to sabotage the outcomes of a distributed Grover's algorithm [86]. The integration of QEC and fault-tolerant strategies is thus crucial [87]. Even though they increase the complexity of quantum operations, they are indispensable for realizing robust quantum computing systems.

Moreover, incorrectly encoding problem instances into quantum states can introduce significant vulnerabilities, particularly in the Variational Quantum Eigensolver (VQE) algorithm [88]. VQE is a hybrid quantum-classical algorithm that uses a parameterized quantum circuit to approximate the lowest eigenvalue of a Hamiltonian. This Hamiltonian represents the energy function of a system, and the algorithm aims to minimize this function to determine the system's ground state energy. It is particularly useful in quantum chemistry and material science. However, misrepresenting the Hamiltonian can lead the algorithm to optimize an incorrect model, resulting in flawed solutions [89]. This misrepresentation could be due to encoding errors or tampering, making the system susceptible to manipulation. Such vulnerabilities highlight the need for rigorous validation and verification in quantum computing to ensure the integrity and accuracy of computational results.

Finally, quantum algorithms and protocols often depend on specific assumptions or approximations that simplify complex quantum challenges. If these assumptions prove incorrect or are misapplied, they can create vulnerabilities. For instance, QKD setups often assume that the equipment operates flawlessly and accurately represents quantum states without unintended defects. However, quantum devices are typically noisy, which can obscure the presence of an eavesdropper. When the end parties compare results, distinguishing between noise and actual eavesdropping can be challenging. Thus, given the fast-paced advancements in this field, it is essential to revisit applications and security strategies to address emerging vulnerabilities consistently.

## D.2. Quantum Probes

In quantum networks that facilitate "prepare and measure" QKD, attackers can exploit vulnerabilities in the transmitted states to extract data [90]. There are three categories of such attacks: individual, collective, and coherent [91]. It is worth noting that while entanglement-based and MDI-QKD can potentially avoid such attacks, they are included here for completeness. This is because "prepare and measure" QKD remains an active area of research, and new attacks evolving from those discussed here continue to be developed [92], [93].

In individual attacks, Eve employs a quantum probe for each qubit sent between Alice and Bob, designing the probe to become entangled with the transmitted qubits. This entanglement allows her to extract information without being detected by Alice and Bob. The critical challenge for Eve is to engineer her quantum probes and their measurements to minimize any disturbance to the qubits, as any significant disruption can increase the Quantum Bit Error Rate (QBER). An increased QBER would alert Alice and Bob to potential eavesdropping. After Alice and Bob disclose their measurement bases (which they use to decode the qubit), Eve measures her entangled probes to glean information about the transmission, potentially compromising or altering the quantum data. While this strategy is commonly associated with QKD, it can threaten the confidentiality of any quantum communication. The risk of these attacks can be mitigated by setting a threshold for acceptable interference or error rates and promptly responding if these thresholds are exceeded.

Collective attacks are more sophisticated than individual ones, involving Eve entangling her probes with blocks of qubits instead of single units. She then treats these entangled qubits as a collective system, enabling her to exploit correlations that may emerge from QEC processes and the exchange of quantum information between Alice and Bob, potentially leading to more substantial data extraction. Employing randomized sequences in quantum transmissions can complicate Eve's data extraction. When combined with advanced QEC codes, these measures can significantly reduce the information that Eve can extract [94].

Coherent attacks involve Eve treating all transmitted qubits as one unified system. A high-dimensional quantum probe entangles the entire quantum state being sent. She then carries out a joint measurement of this entangled state. It allows Eve to access all transmitted data simultaneously. To counteract these attacks, sending extra states or photons at various intensities makes it challenging for Eve to determine which states carry real information [41]. Frequently changing the patterns of quantum transmission can also disrupt coherent attacks, making it harder to treat the entire quantum transmission as a single, coherent system.

TABLE II
READINESS OF QUANTUM ATTACKS

| Layers | Attacks | Requirements | Readiness |
|---|---|---|---|
| Application Layer | Quantum Algorithmic Attacks | Access to quantum systems, access to classical controllers | High – Feasible with the algorithm or protocol knowledge and access to classical systems. |
| Application Layer | Quantum Probes | Devices to interact with quantum systems and measure or store states | Low – Demands advanced equipment and expertise. |
| Network Layer | Untrusted Repeater Nodes | Control or compromise of network nodes | Low to Moderate – Exploits network trust, requiring access to intermediate nodes. |
| Network Layer | DoS | Classical computing resources to flood requests | High – Relatively simple to execute using classical tools to overwhelm the network. |
| Network Layer | Routing Disruption | Access to classical network systems, tools for intercepting classical data | High – Feasible with protocol knowledge and access to classical control systems. |
| Link Layer | Entangling-probe | Access to quantum channels, entanglement sources | Low – Requires access to endpoints and advanced hardware to generate entanglement. |
| Link Layer | MitM | Access to quantum channels, noise injection devices | Moderate – Feasible with access but challenging to execute without detection. |
| Link Layer | Attacks on QEC | Access to quantum channels, noise injection devices | Moderate – Requires access to endpoints. |
| Physical Layer | PNS | Quantum non-demolition devices, quantum memory, precise photon detectors | Low - Requires advanced quantum-specific hardware and precise manipulation of photon states. |
| Physical Layer | Trojan-Horse | Bright light sources, optical couplers, precise photon detectors | Moderate - Some components are readily available. |
| Physical Layer | Phase-Remapping | Phase modulators, polarization controllers | Low - Demands precise control over phase settings. |

## VI. DISCUSSION – READINESS OF QUANTUM ATTACKS

The complexity of executing a quantum attack depends on various factors, such as technical requirements, hardware accessibility, and available computational resources. This section explores these aspects for the discussed attacks, emphasizing their feasibility and potential impact.

Most attacks on the physical layer, such as PNS or Trojan-Horse attacks, require access to specialized quantum devices, including precise photon detectors, quantum memories, and sources capable of generating specific quantum states. These tools are not widely available and often require significant expertise and controlled environments. While certain components for some attacks, such as bright light sources used in Trojan-Horse attacks, can be commercially obtained, executing them typically necessitates additional equipment—such as optical couplers, connectors, and precise detectors—that are not easily accessible. Additional attacks, such as manipulating gated avalanche photodiodes and phase-remapping, involve sophisticated control over quantum devices and their phase settings, requiring high precision and expertise. Moreover, attacks in this layer must be performed in real time, as quantum states are highly sensitive to environmental changes. Therefore, while feasible, these attacks are generally challenging for an outsider to execute.

At the link layer, direct quantum channels between nodes are potential targets for disruption. Such attacks are more resource-efficient, but similarly, there are simpler strategies to counter them. For example, injecting noise into the channel to compromise transmission does not necessitate additional equipment once access is gained for MitM attacks. Meanwhile, such attacks can be detected by employing an interlock protocol, as described in Subsection IV.B.2. More complex attacks, such as side-channel exploitation during link initialization and targeted attacks on error correction protocols, leverage implementation flaws. These attacks are considerably more challenging, requiring high precision and advanced tools.

Attacks on the network layer, particularly those targeting hybrid quantum-classical systems, demand significant classical computing power. For instance, DoS attacks on quantum routing or synchronization rely on classical resources to flood the system with a high volume of requests or disruptions. In contrast, DoS attacks targeting QKD networks can be executed by gaining access to the QKD channel, which requires quantum apparatus rather than classical computing power. Alternatively, attackers may use classical computers to simulate quantum systems, gain insights, or prepare for interventions. However, the computational resources required for such simulations increase rapidly with the complexity of the quantum system.

Application-layer attacks, which focus on tasks including key exchange or computations, are often the most resource-efficient for adversaries. These attacks typically exploit vulnerabilities in protocols, algorithms, or system architectures and generally do not require additional hardware. For instance, targeting distributed quantum algorithms, such as the collaborative execution of Shor's or Grover's algorithms, involves exploiting protocol weaknesses, such as timing or consensus issues in distributed architectures. However, quantum probes targeted at QKD systems, which involve direct interaction with quantum systems to extract or disrupt information, require hardware and expertise, making them more challenging to execute.

Overall, executing attacks on the quantum internet requires varying levels of resources and expertise, depending on the target and method. While higher-layer vulnerabilities can be exploited with relatively accessible resources compared to lower-layer attacks, most still demand specialized hardware and significant computing power. For ease of reference, we have compiled the readiness of the discussed attacks in Table II.

## VII. CONCLUSION

The quantum internet offers enhanced communication and distributed quantum computing opportunities, yet it also poses unique security challenges inherent to quantum mechanics. These challenges span various layers, including physical, link, network, and application. A thorough, multi-layered security investigation that addresses concerns from various perspectives is essential. This paper conducts a layer-wise security analysis to stimulate further research into quantum internet security, paving the way for a secure and advanced future internet.

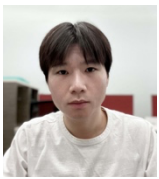

**Zebo Yang** (Graduate Student Member, IEEE) received an M.E. in computer engineering from Waseda University, Japan (2019). He is currently pursuing a Ph.D. in computer science at Washington University in St. Louis, USA. From 2011 to 2017, he worked as a Software Engineer at Tencent, Baidu, and DJI. Since 2019, he has been a Research Assistant at Washington University. His research interests include quantum computing, quantum networks, networking, network security, and machine learning.

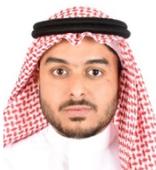

**Ali Ghubaish** (Student Member, IEEE) received a B.S. (Hons.) in computer engineering with a networking minor from Prince Sattam Bin Abdulaziz University, Saudi Arabia, in 2013. He obtained an M.S. and a Ph.D. in computer engineering from Washington University in St. Louis, USA, in 2017 and 2024, respectively. He is currently a lecturer at Prince Sattam Bin Abdulaziz University in Al-Kharj, Saudi Arabia. Ali has been a Graduate Research Assistant at Washington University since 2018. His research focuses on network and system security, IoT, the Internet of Medical Things, and healthcare systems.

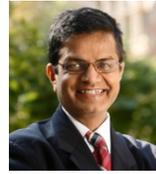

**Raj Jain** (Life Fellow, IEEE) received a B.E. in electrical engineering from APS University, India, and an M.E. from the Indian Institute of Science. He obtained his Ph.D. in Applied Maths (computer science) from Harvard University. He is the Barbara J. and Jerome R. Cox, Jr., Professor at Washington University in St. Louis. Dr. Jain co-founded Nayna Networks, Inc., and has held significant roles in academia and industry. He has received the 2018 James B. Eads Award and the 2017 ACM SIGCOMM Life-Time Achievement Award. He is known as one of the most cited authors in computer science and has authored "The Art of Computer Systems Performance Analysis." He is a Fellow of IEEE, ACM, and AAAS.

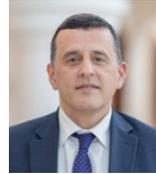

**Ala Al-Fuqaha** (Senior Member, IEEE) received his Ph.D. in Computer Engineering and Networking from the University of Missouri-Kansas City, Kansas City, MO, USA. He is a professor at the Information and Computing Technology division, College of Science and Engineering, Hamad Bin Khalifa University (HBKU). He is a senior member of the IEEE, a senior member of the ACM, and an ABET Program Evaluator (PEV) and commissioner. He serves on editorial boards of multiple journals, including IEEE Communications Letters, IEEE Network Magazine, and Springer AJSE.

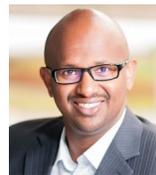

**Aiman Erbad** (Senior Member, IEEE) received a B.Sc. in computer engineering from the University of Washington (2004), a master's degree in embedded systems and robotics from the University of Essex (2005), and a Ph.D. in computer science from The University of British Columbia (2012). He is a Professor of Computer Engineering and the VP of Research and Graduate Studies at Qatar University. His research areas include cloud and edge computing, edge intelligence, network intelligence, IoT, and Blockchains. Dr. Erbad is also an editor and program chair for various journals and conferences and a Senior Member of ACM.

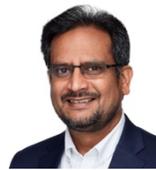

**Ramana Kompella** (Member, IEEE) received a B.Tech. from the Indian Institute of Technology, Bombay (1999), a M.S. from Stanford University (2001), and a Ph.D. from the University of California, San Diego (2007), all in computer science and engineering. He serves as a Distinguished Engineer and the Head of Research in Cisco's Emerging Tech and Incubation group, leading university research collaborations. His research focuses on data centers, cloud performance optimization, and router algorithmics. Dr. Kompella has been an ACM member since 2007, receiving the NSF CAREER Award in 2011, and has Best Paper awards at conferences like ACM SOCC.

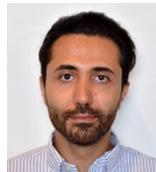

**Hassan Shapourian** (Member, IEEE) received a M.S. in electrical engineering from Princeton University (2013). He has a Ph.D. in Theoretical Physics from the University of Chicago (2019) and worked as a postdoctoral researcher at MIT/Harvard and Microsoft Station Q. He is a Senior Quantum Researcher at Cisco, leading projects on quantum information processing and hardware physics. He is a recipient of Microsoft Research Postdoctoral Fellowship, Simons Postdoctoral Fellowship, Kavli Institute for Theoretical Physics (KITP) Graduate Fellowship, Biruni Graduate Student Research Award, John Bardeen Award (by UIUC), and Princeton University Fellowship.

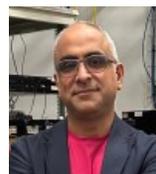

**Reza Nejabati** (Member, IEEE) is the Head of Quantum Research and Quantum Lab at Cisco, focused on advancing quantum networking. Previously, he was Chair Professor of Networks and Head of the High-Performance Network Group at the University of Bristol, UK. He is also a visiting professor and Cisco Chair at Curtin University's Cisco Centre for Intent-Based Networking in Australia. Reza established globally recognized research in autonomous and quantum networks, co-founded Zeetta Networks Ltd, and received the IEEE Charles Kao Award in 2016 for contributions to 5G, smart cities, quantum communication, and future Internet experimentation.